\documentclass{article}
\usepackage{spconf,amsmath,graphicx,hyperref}

\usepackage{graphicx}

\usepackage{enumitem}
\usepackage{booktabs}  
\usepackage{multirow}
\usepackage{pifont}
\usepackage{xspace}
\usepackage{listings}
\usepackage{xcolor}
\usepackage{colortbl}
\usepackage{nicematrix}
\usepackage{afterpage}
\usepackage{textcomp}

\usepackage{algorithm}
\usepackage[noend]{algpseudocode}
\usepackage{amsmath}
\usepackage{bbm}
\usepackage{comment}

\usepackage{amssymb}

\newcommand\ours{SSEmb\xspace}
\definecolor{upperbound}{gray}{0.90}

\title{SSEmb: Training a Generalizable Speech Retrieval Model Using Only
Automatic Speech Recognition Data}
\name{Sangkwon Park$^{1}$, Dongwook Lee$^{2}$, Youngho Cho$^{2}$, Donghun Kang$^{1}$,  Heeseung Kim$^{4\dagger}$, Sungroh Yoon$^{1,2,3\dagger}$\thanks{$\dagger$ Corresponding Authors}}
\address{$^{1}$ECE, Seoul National University\quad$^{2}$IPAI, Seoul National University\\ $^{3}$AIIS, ASRI, INMC, and ISRC, Seoul National University\quad$^{4}$University of Seoul}

\begin{document}
%
\maketitle
\begin{abstract}
End-to-end speech-to-text retrieval is becoming an important capability for voice agents, allowing spoken queries to retrieve relevant documents without intermediate transcription. Training these retrievers typically depends on large collections of spoken query-document pairs that are costly to acquire and difficult to extend across domains and languages. We introduce \ours, which replaces such task-specific supervision with speech-transcript pairs from standard ASR corpora. A lightweight adapter projects the output of a frozen speech encoder into the retrieval space of a frozen pretrained text embedding model. \ours is competitive with cascaded ASR-then-retrieve systems while reducing inference latency, and outperforms existing publicly available speech embedding models. It also transfers to text-to-speech and speech-to-speech retrieval without additional training. Controlled analyses validate the speech-transcript alignment paradigm, demonstrating robustness to reduced training data and consistent gains from larger corpora. Further ablations of loss functions and trainable modules establish practical guidelines for the framework. Model weights and codes are to be released.
\end{abstract}
\section{Introduction}
\label{sec}

Modern voice agents ~\cite{xu2025qwen3omni} increasingly rely on external knowledge to answer spoken requests, making speech-driven retrieval a key component of natural spoken interaction~\cite{gao2025lucy}. Conventional systems first transcribe a spoken query using Automatic Speech Recognition (ASR) and then pass the transcript to a text retriever~\cite{spokencontentretrieval}. Although this cascade benefits from powerful text retrieval models, sequential ASR decoding adds latency, and transcription errors propagate directly to retrieval~\cite{speechrag,chen2025wavrag}. End-to-end speech-to-text retrieval instead encodes speech directly as a retrieval query, avoiding explicit transcription~\cite{lin2024speechdpr,sun2025seal}.

Training such end-to-end retrievers, however, generally relies on paired spoken queries and relevant documents, which are costly to collect and difficult to scale across domains and languages. SpeechRAG demonstrated that such retrieval-specific supervision can be avoided by aligning speech with its transcript embedding, transferring the retrieval capability of a pretrained text embedding model to speech through speech-transcript pairs~\cite{speechrag}. While this result establishes the potential of speech-transcript alignment, its broader applicability as a training paradigm for speech retrieval has remained largely unexplored. We extend this idea by using standard ASR corpora as the sole source of supervision, turning widely available speech-transcript data into a scalable resource for training speech retrievers without retrieval-specific annotation. We then systematically examine its scalability and generalizability across data and model scales, languages, retrieval directions, and training configurations.


\begin{figure}[!t]
\centering
\centerline{\includegraphics[width=8.5cm]{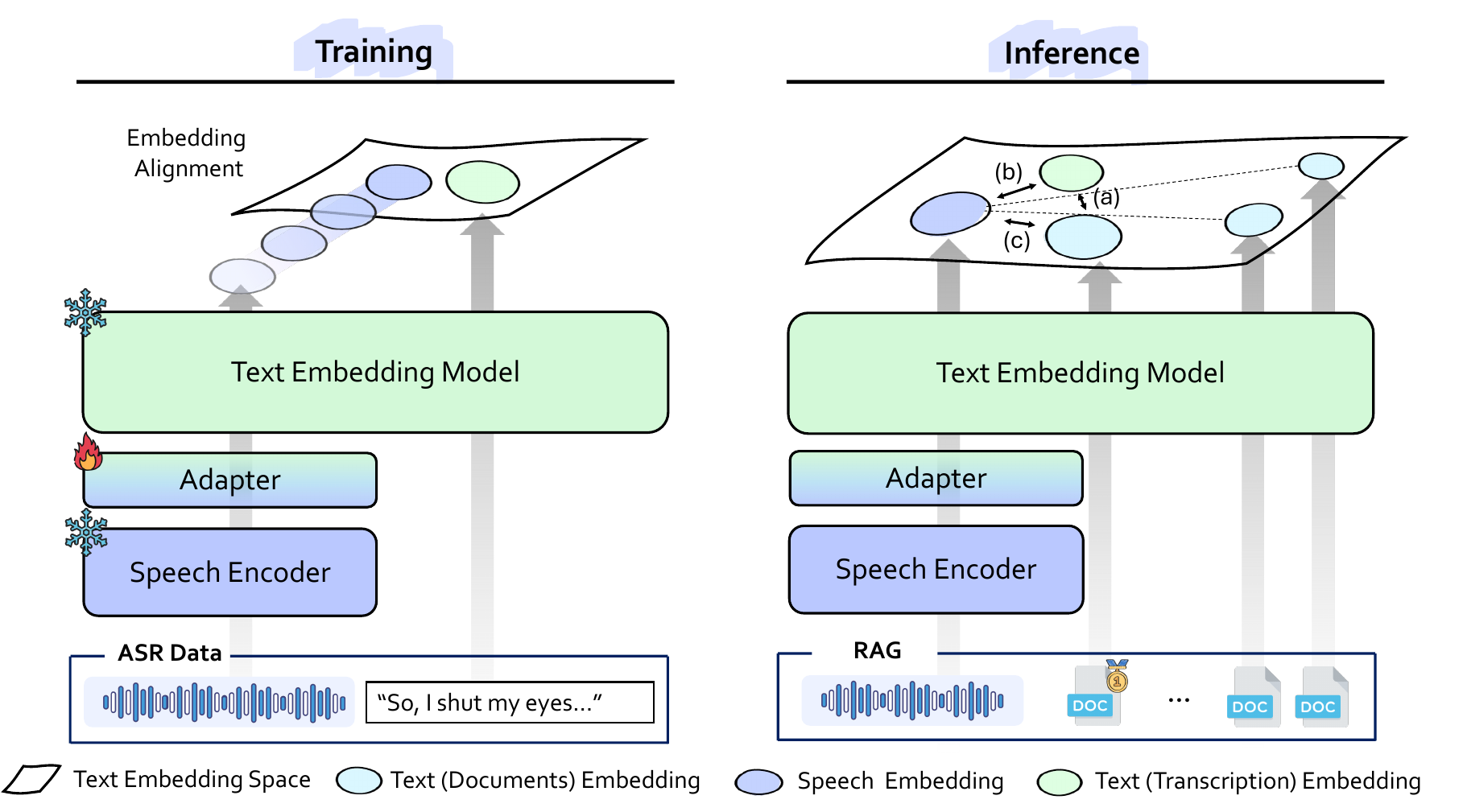}}
\caption{Overview of the \ours architecture. The acoustic representation of speech passes through an adapter into the text embedding model to produce speech embeddings. ASR data is used to align the embeddings of speech with those of its transcript. At inference, the speech query is embedded in the same manner for retrieval.}
\label{fig:architecture}
\end{figure}

\begin{table*}[ht]
\centering
\caption{Retrieval performance across three English benchmarks. Shaded rows indicate cascaded systems corresponding to each model, where the transcript of the input query is directly used for retrieval. $\Delta t$ denotes the latency of each method, measured in seconds. Speedup over the corresponding cascaded baseline is also reported.}
\resizebox{\textwidth}{!}{
\begin{tabular}{l ccc ccc ccc c}
\toprule
 & \multicolumn{3}{c}{\textbf{HotpotQA}} & \multicolumn{3}{c}{\textbf{MuSiQue}} & \multicolumn{3}{c}{\textbf{Mr.TyDi (Eng)}} & \\
\cmidrule(lr){2-4} \cmidrule(lr){5-7} \cmidrule(lr){8-10}
\textbf{Model} & R@2 & R@5 & nDCG@10 & R@2 & R@5 & nDCG@10 & R@2 & R@5 & nDCG@10 & \textbf{$\Delta t$} (speedup) \\
\midrule
\rowcolor{upperbound} Cascaded Q3E-0.6B & 0.4940 & 0.6942 & 0.7267 & 0.4535 & 0.6614 & 0.7201 & 0.6453 & 0.8356 & 0.6976 & 0.348 \\
\textbf{\ours (Q3E-0.6B)} & 0.4736 & 0.6729 & 0.7049 & 0.4162 & 0.6141 & 0.6764 & 0.5703 & 0.7860 & 0.6450 & 0.144 (\textbf{2.42$\times$)} \\
\addlinespace
\rowcolor{upperbound} Cascaded Q3E-4B & 0.5369 & 0.7553 & 0.7772 & 0.4839 & 0.6987 & 0.7494 & 0.7048 & 0.8774 & 0.7429 & 0.386 \\
\textbf{\ours (Q3E-4B)} & 0.5215 & 0.7373 & 0.7609 & 0.4353 & 0.6484 & 0.6998 & 0.6659 & 0.8495 & 0.7135 & 0.145 (\textbf{2.66$\times$)} \\
\addlinespace
\rowcolor{upperbound} Cascaded Q3E-8B & 0.5425 & 0.7656 & 0.7886 & 0.5068 & 0.7277 & 0.7788 & 0.7389 & 0.8875 & 0.7616 & 0.405 \\
\textbf{\ours (Q3E-8B)} & 0.5194 & 0.7380 & 0.7656 & 0.4585 & 0.6767 & 0.7280 & 0.6961 & 0.8675 & 0.7238 & 0.151 (\textbf{2.68$\times$)} \\
\midrule
SONAR & 0.2877 & 0.4726 & 0.5009 & 0.2397 & 0.4260 & 0.4757 & 0.1035 & 0.2396 & 0.2101 & 0.083 \\
\bottomrule
\end{tabular}
}
\label{tab:main_en}
\end{table*}

We introduce \textbf{\ours} (\textbf{S}imple \textbf{S}peech \textbf{Emb}edding), a modular end-to-end speech retriever trained using only standard ASR corpora. \ours connects a frozen speech encoder to a frozen text embedding model through a lightweight adapter. We train the adapter with cosine alignment and InfoNCE losses to align the resulting speech embedding with the text model's embedding of the corresponding transcript, thereby transferring the embedding space of the pretrained text model to speech without requiring spoken query-document pairs. In this way, a spoken query can be embedded directly for retrieval without ASR.

We evaluate \ours with text embedding models ranging from 0.6B to 8B parameters on English and six additional languages. \ours approaches the retrieval performance of cascaded ASR-then-retrieve systems with substantially lower latency, outperforms existing speech embedding model, and surpasses the cascaded system in some low-resource settings. It also generalizes to text-to-speech and speech-to-speech retrieval without additional training. Controlled analyses validate the speech-transcript alignment paradigm across data scales, showing only modest performance degradation when training data is reduced to 10 hours and consistent improvements as more data becomes available. Together with further ablations on loss functions and trainable modules, these findings provide a broader understanding of the framework and establish practical guidelines for effectively transferring pretrained text retrieval capabilities to speech.

\begin{table*}[htbp]
\centering

\caption{Retrieval performance of \ours on Mr.TyDi across six low-resource languages.}
\begin{tabular}{l ccc ccc ccc}
\toprule
 & \multicolumn{3}{c}{\textbf{Korean (960h)}} & \multicolumn{3}{c}{\textbf{Japanese (315h)}} & \multicolumn{3}{c}{\textbf{Thai (40h)}} \\
\cmidrule(lr){2-4} \cmidrule(lr){5-7} \cmidrule(lr){8-10}
\textbf{Model} & R@2 & R@5 & nDCG@10 & R@2 & R@5 & nDCG@10 & R@2 & R@5 & nDCG@10 \\
\midrule
\rowcolor{upperbound} Cascaded & 0.6715 & 0.8265 & 0.7097 & 0.6430 & 0.8380 & 0.6816 & 0.6387 & 0.8037 & 0.6934 \\
\textbf{\ours} & 0.5805 & 0.7805 & 0.6568 & 0.5210 & 0.7310 & 0.5864 & 0.6615 & 0.8468 & 0.7087 \\
 
SONAR & 0.1770 & 0.3470 & 0.2975 & 0.1360 & 0.3090 & 0.2607 & 0.1403 & 0.3018 & 0.2567 \\
\midrule
 & \multicolumn{3}{c}{\textbf{Indonesian (35h)}} & \multicolumn{3}{c}{\textbf{Russian (55h)}} & \multicolumn{3}{c}{\textbf{Arabic (80h)}} \\
\cmidrule(lr){2-4} \cmidrule(lr){5-7} \cmidrule(lr){8-10}
\textbf{Model} & R@2 & R@5 & nDCG@10 & R@2 & R@5 & nDCG@10 & R@2 & R@5 & nDCG@10 \\
\midrule
\rowcolor{upperbound} Cascaded & 0.6560 & 0.8300 & 0.7150 & 0.6610 & 0.8220 & 0.7010 & 0.7240 & 0.8870 & 0.7520 \\
\textbf{\ours} & 0.5950 & 0.7970 & 0.6643 & 0.5840 & 0.7810 & 0.6384 & 0.6260 & 0.8230 & 0.6729 \\
SONAR & 0.1000 & 0.2540 & 0.2136 & 0.1690 & 0.3200 & 0.2705 & 0.1110 & 0.2730 & 0.2355 \\
\bottomrule
\end{tabular}

\label{tab:main_multilingual}
\vskip -0.05in
\end{table*}

\section{SSEmb}
\label{sec:ssemb}

\subsection{Architecture}

Following prior adapter-based approaches~\cite{speechrag}, we construct a
speech retriever by connecting a pretrained speech encoder to a frozen text
embedding model through a lightweight adapter.

We use Qwen3-Embedding~\cite{zhang2025qwen3embedding} as the text backbone,
denoted by $f_{\mathrm{text}}$. Given a text input $x$,  representing either a query $q$ or a passage $c$, we obtain its $d$-dimensional embedding by pooling the hidden state of an appended EOS token:
\begin{equation}
\mathbf{z}_x =
\left[
f_{\mathrm{text}}
\left(E(x) \oplus E_{\texttt{[EOS]}}\right)
\right]_{\texttt{[EOS]}},
\qquad x \in \{q,c\},
\end{equation}
where $E(x)$ denotes the sequence of token embeddings and $\oplus$ denotes
sequence concatenation. Retrieval scores are computed using cosine similarity of the embeddings.

Given a speech input $a$, the speech encoder $f_{\mathrm{speech}}$ produces a sequence of $d_s$-dimensional acoustic representations, which the adapter $g_{\theta}$ projects into the input space of the text model. The resulting representations are treated as pseudo-token embeddings and processed by the frozen text model:
\begin{align}
\mathbf{H}_a &= f_{\mathrm{speech}}(a)
    \in \mathbb{R}^{T \times d_s}, \\
\widetilde{\mathbf{E}}_a &= g_\theta(\mathbf{H}_a)
    \in \mathbb{R}^{T \times d}, \\
\mathbf{z}_a &=
\left[
f_{\mathrm{text}}
\left(
\widetilde{\mathbf{E}}_a
\oplus E_{\texttt{[EOS]}}
\right)
\right]_{\texttt{[EOS]}}
\in \mathbb{R}^{d}.
\end{align}

The resulting $\mathbf{z}_a$ serves as the final speech embedding. The architecture of $g_{\theta}$ varies by language, as described in Section \ref{ssec:multiple}.
Following Qwen3-Embedding, an instruction is prepended to text queries but not to passages. We omit this instruction from the notation for brevity.

\subsection{Training Objective}

Given a batch of speech-transcript pairs
$\{(a_i,q_i)\}_{i=1}^{B}$ from standard ASR corpora, we compute the transcript
embeddings using the frozen text model and align the corresponding speech
embeddings with them.  Let $s_{ij} =
\mathbf{z}_{a_i}^\top \mathbf{z}_{q_j} /
(\lVert\mathbf{z}_{a_i}\rVert
\lVert\mathbf{z}_{q_j}\rVert)$ denote their cosine similarity. The cosine alignment loss is
\begin{equation}
\mathcal{L}_{\mathrm{cos}} =
\frac{1}{B}\sum_{i=1}^{B}\left(1-s_{ii}\right).
\end{equation}
We additionally use an InfoNCE loss with in-batch
transcripts as negatives:
\begin{equation}
\mathcal{L}_{\mathrm{InfoNCE}} =
-\frac{1}{B}\sum_{i=1}^{B}
\log
\frac{\exp(s_{ii}/\tau)}
{\sum_{j=1}^{B}\exp(s_{ij}/\tau)},
\end{equation}
where $\tau$ is the temperature. The final objective is
\begin{equation}
\mathcal{L} =
\mathcal{L}_{\mathrm{cos}}
+\lambda\mathcal{L}_{\mathrm{InfoNCE}},
\end{equation}
where $\lambda$ controls the contribution of the InfoNCE term.

\subsection{Extension to Multiple Languages}
\label{ssec:multiple}
Because both the speech encoder and text embedding model support multiple languages, \ours can be adapted to a new language by training an adapter $g_{\theta}$ on ASR data in that language. For English, we use a two-layer MLP comprising a projection from $d_s$ to $d$, a GELU activation, a second $d$-dimensional projection, and layer normalization. For lower-resource languages, where limited supervision makes it particularly important to exploit the pretrained encoder representations effectively, we use a deeper convolutional adapter with $N=3$ residual blocks. Each block combines a causal 1-D convolution for capturing local temporal patterns with a feed-forward network consisting of two linear projections and a GELU activation, with dropout applied within the block.

\section{EXPERIMENTS}
\label{sec:typestyle}

We use the Audio Transformer encoder from Qwen3-Omni-30B-A3B-Instruct~\cite{xu2025qwen3omni} as the speech encoder. For the text retriever, we adopt Qwen3-Embedding~\cite{zhang2025qwen3embedding} at three model scales: 0.6B, 4B, and 8B parameters. Both the audio encoder and the text retriever are frozen during training. For English, \ours is trained on the 190-hour clean subset of LibriTTS~\cite{zen2019libritts}. For low-resource languages, we use subsets of Common Voice~\cite{ardila2020commonvoice} and KsponSpeech~\cite{bang2020kspon}.

We evaluate speech embedding quality using recall and nDCG@10 on speech-to-text retrieval. As no established benchmark exists for this task, we convert queries from textual retrieval benchmarks into speech using an existing TTS model ~\cite{chatterboxtts2025}. Although synthesized queries may not fully capture the variability of natural speech, they provide a controlled and reproducible evaluation protocol~\cite{chen2025wavrag}. For English, we use HotpotQA~\cite{yang2018hotpotqa}, MuSiQue~\cite{trivedi2022musique}, and the English subset of Mr.~TyDi~\cite{zhang2021mrtydi}; for other languages, we use the corresponding Mr.~TyDi subsets. We additionally evaluate spoken-document retrieval on SpokenSQuAD~\cite{li2018spoken} in both Text-to-Speech and Speech-to-Speech settings.

We compare \ours with a cascaded system that first transcribes speech queries using \textit{whisper-large-v3-turbo}~\cite{radford2023whisper} and then performs retrieval using the same Qwen3-Embedding model. We also compare against SONAR~\cite{Duquenne:2023:sonar_arxiv}, a multilingual embedding model 
trained for text-to-speech translation that has also demonstrated effectiveness in retrieval~\cite{kim-etal-2025-voice-assistant}.

\section{RESULTS and analysis}
\label{sec:majhead}

\subsection{Main Results}

Table~\ref{tab:main_en} presents the results on English benchmarks of \ours with Qwen3-Embedding as the backbone at three model scales. \ours consistently retains the majority of the cascaded baseline's performance across all benchmarks and model sizes. \ours also surpasses SONAR by a wide margin on all benchmarks, highlighting the advantage of aligning to a retrieval-specialized embedding space over using a general-purpose cross-modal representation. \ours consistently shows lower latency than its cascaded counterparts, showing the effectiveness of end-to-end speech retrieval.

Table~\ref{tab:main_multilingual} presents the results of \ours across six languages (Korean, Japanese, Thai, Indonesian, Russian, and Arabic) using Qwen3-Embedding-0.6B as the base model. \ours achieves performance comparable to the cascaded pipeline without any retrieval-specific supervision, demonstrating that speech-transcript alignment generalizes beyond English. Notably, \ours surpasses the cascaded baseline in Thai, likely because higher ASR error rates propagate transcription errors into retrieval, highlighting the advantage of end-to-end speech retrieval when ASR is less reliable.

\begin{table}[t]
\setlength{\tabcolsep}{3pt}
\renewcommand{\arraystretch}{1.15}
\centering
\small

\caption{Retrieval performance on SpokenSquad comparison across Text2Speech and Speech2Speech settings. WavRAG$^{\dagger}$ results are taken directly from the original paper \cite{chen2025wavrag}.}
\begin{tabular*}{\columnwidth}{@{\extracolsep{\fill}} l cccc @{}}
\toprule
& R@1 & R@5 & R@10 & nDCG@10 \\
\midrule
\multicolumn{5}{@{}l}{\textbf{Text2Speech}} \\
\quad Cascaded & 0.6978 & 0.9086 & 0.9507 & 0.8276 \\
\quad \ours & \textbf{0.6997} & \textbf{0.9168} & \textbf{0.9632} & 0.8342 \\
\quad WavRAG$^{\dagger}$ & 0.6844 & 0.8374 & 0.9023 & \textbf{0.8483} \\
\midrule
\multicolumn{5}{@{}l}{\textbf{Speech2Speech}} \\
\quad Cascaded & 0.6870 & 0.9047 & 0.9497 & 0.8218 \\
\quad \ours & 0.6840 & 0.9036 & 0.9510 & 0.8210 \\
\bottomrule
\end{tabular*}
\label{tab:main_spokendoc}
\end{table}

Table~\ref{tab:main_spokendoc} shows the performance of \ours on spoken document settings. In the Text-to-Speech setting, \ours outperforms the cascaded baseline across most metrics and achieves performance comparable to WavRAG, despite using a smaller model. In Speech-to-Speech setting, \ours closely matches the cascaded baseline, demonstrating that the adapter-based alignment transfers effectively to spoken document encoding without additional training.

\subsection{The Speech-Transcript Alignment Paradigm}
\label{ssec:analysis_paradigm}

We compare four training configurations to evaluate speech–\allowbreak transcript alignment against explicit speech–document supervision. Our default model uses LibriTTS speech-transcript pairs and the combined loss (\textit{Ours} in Table~\ref{tab:ablation_all}). To separate the effects of supervision and data domain, we additionally train on synthesized HotpotQA spoken queries~\cite{chatterboxtts2025} using either speech-transcript alignment (\textit{w/ ASR}) or speech-document pairs with the combined loss (\textit{w/ Q-D}) and InfoNCE alone (\textit{w/ Q-D \& I.}). The results show that speech-transcript alignment achieves the best retrieval performance, outperforming direct speech-document supervision. 
The frozen text embedding model already captures query–document relationships, allowing \ours to inherit this retrieval capability by directly matching speech to its transcript embedding. In contrast, spoken query–document training must bridge the modality gap indirectly through document-level relevance signals, making the learning objective more complex and difficult to optimize.
\subsection{Dataset Scale}
Language-specific adapters trained on ASR corpora of varying sizes achieve reasonable retrieval quality. To examine this relationship systematically, we vary the amount of English ASR training data and measure its effect on retrieval quality. As shown in Table~\ref{tab:ablation_all}, performance degrades gracefully as training data is reduced: down to approximately 10 hours, retrieval quality remains within a reasonable margin of the full setup, demonstrating that the alignment framework is data-efficient. Additionally, incorporating over 46k hours of diverse data from the Emilia dataset~\cite{he2024emilia} further improves retrieval performance. This indicates that the framework can effectively leverage large-scale ASR resources when available.

\begin{table}[t]
\setlength{\tabcolsep}{3pt}
\renewcommand{\arraystretch}{1.15}
\centering
\small

\caption{Ablation study on training strategies using the 0.6B model. We compare different training paradigms, the amount of training data, loss, and trained modules. Cascaded denotes the upper-bound pipeline. nDCG refers to nDCG@10.}
\resizebox{\columnwidth}{!}{%
\begin{tabular}{l ccc ccc}
\toprule
& \multicolumn{3}{c}{\textbf{HotpotQA}} & \multicolumn{3}{c}{\textbf{MuSiQue}} \\
\cmidrule(lr){2-4} \cmidrule(lr){5-7}
& R@2 & R@5 & nDCG & R@2 & R@5 & nDCG \\
\midrule
\rowcolor{upperbound} Cascaded & 0.4940 & 0.6942 & 0.7267 & 0.4535 & 0.6614 & 0.7201 \\
Ours & 0.4736 & 0.6729 & 0.7049 & 0.4162 & 0.6141 & 0.6764 \\
\midrule\midrule
\textit{Training paradigm} \\
\quad w/ ASR & \textbf{0.4936} & \textbf{0.6943} & \textbf{0.7240} & \textbf{0.4253} & \textbf{0.6286} & \textbf{0.6921} \\
\quad w/ Q-D & 0.4593 & 0.6627 & 0.6879 & 0.3796 & 0.5915 & 0.6475 \\
\quad w/ Q-D \& I. & 0.4710 & 0.6739 & 0.7014 & 0.3994 & 0.6044 & 0.6646 \\
\midrule\midrule
\textit{Training scale} \\
\quad w/ 50h & 0.4624 & 0.6649 & 0.6957 & 0.4079 & 0.6030 & 0.6667 \\
\quad w/ 10h & 0.4411 & 0.6406 & 0.6710 & 0.3887 & 0.5941 & 0.6498 \\
\quad w/ 2h & 0.3842 & 0.5746 & 0.6069 & 0.3496 & 0.5446 & 0.6076 \\
\quad w/ Emilia (46k h) & 0.\textbf{4906} & \textbf{0.6925} & \textbf{0.7231 }& \textbf{0.4324} & \textbf{0.6245 }& \textbf{0.6869} \\
\midrule\midrule
\textit{Loss selection} \\
\quad Cosine & 0.4662 & 0.6660 & 0.6976 & 0.4149 & 0.6138 & 0.6742 \\
\quad InfoNCE & 0.4640 & 0.6676 & 0.6976 & 0.4008 & 0.5997 & 0.6561 \\
\midrule\midrule
\textit{Trained modules} \\
\quad Embedding & 0.0561 & 0.1390 & 0.1591 & 0.0977 & 0.2405 & 0.2905 \\
\quad Encoder & 0.0419 & 0.1126 & 0.1389 & 0.1548 & 0.3272 & 0.3827 \\
\quad Both & 0.0623 & 0.1462 & 0.1708 & 0.0947 & 0.2433 & 0.2982 \\
\bottomrule
\end{tabular}%
}
\label{tab:ablation_all}
\end{table}

\subsection{Loss Ablation and Trainable Modules}

Table~\ref{tab:ablation_all} compares cosine alignment, InfoNCE, and their combination. The combined objective yields a small but consistent improvement across both benchmarks, indicating that pairwise alignment of cosine loss and contrastive discrimination of InfoNCE loss provide complementary signals.

By default, \ours trains only the adapter while freezing the speech encoder and text embedding model, unlike conventional approaches that fine-tune an embedding backbone~\cite{lin2024speechdpr}. Unfreezing either backbone substantially degrades performance, suggesting that preserving their pretrained acoustic representations and retrieval structure is critical, with the adapter serving as the sole bridge between them.

\section{Conclusion}
\label{sec}

We introduced \ours, an end-to-end speech retriever trained only on standard ASR data by aligning speech with a frozen text retriever's embedding space. Across English and six other languages, \ours approaches cascaded ASR-then-retrieve performance with lower latency and generalizes to text-to-speech and speech-to-speech retrieval.
Our analyses show that speech-transcript alignment outperforms direct speech-document supervision, while retrieval performance consistently improves as the amount of ASR training data increases. These results demonstrate a practical, scalable approach to speech retrieval without task-specific query-document pairs.


\section{Acknowledgments}
This work was supported by the Institute of Information \& Communications Technology Planning \& Evaluation (IITP) grants funded by the Korea government (MSIT) [NO.RS-2021-II211343, Artificial Intelligence Graduate
School Program (Seoul National University); No.2022-0-00959, RS-2022-II220959], by the National Research Foundation of Korea (NRF) grant [No.2022R1A3B1077720, 2022R1A5A7083908], BK21 FOUR Program of the Education and Research Program for Future ICT Pioneers, Seoul National University in 2026. Generative AI tools (Claude and ChatGPT) were used for coding assistance, proofreading and revision of author-written drafts throughout the paper.

{\fontsize{9}{9}\selectfont
\bibliographystyle{IEEEbib}
\bibliography{strings,refs}
}

\end{document}